\documentclass[aps,pra,twocolumn,notitlepage,nofootinbib,superscriptaddress]{revtex4-1}

\usepackage{rahulbhadani_arxiv}
\usepackage{amsmath}
\usepackage{amssymb}
\usepackage{graphicx}
\usepackage[colorlinks=true,allcolors=blue]{hyperref}
\usepackage{braket}
\usepackage{dsfont}
\usepackage{multirow}
\usepackage{subcaption}
\usepackage{listings}
\usepackage{physics}
\usepackage{romanbar}
\newcounter{algorithm}

\newcounter{alg}
\renewcommand{\thealg}{\arabic{alg}}

\begin{document}

\title{Neural-network Generated Quantum State Can Mitigate the Barren Plateau in Variational Quantum Circuits}

\author{Zhehao Yi}
\affiliation{AI, Autonomy, Resilience, Control (AARC) Lab, Electrical \& Computer Engineering, The University of Alabama in Huntsville, AL, 35899, USA}

\author{Rahul Bhadani}
\affiliation{AI, Autonomy, Resilience, Control (AARC) Lab, Electrical \& Computer Engineering, The University of Alabama in Huntsville, AL, 35899, USA}

\date{\today}

\begin{abstract}
We find that using neural networks to generate quantum states can effectively alleviate the barren plateau phenomenon present in random variational quantum circuits.
\end{abstract}
\maketitle

\footnotesize{Paper accepted and presented at Optica Frontiers in Optics $+$ Laser Science 2025 in Denver, Colorado, USA}

\section{Construction of random quantum circuit and NGQS}
\begin{figure*}[htbp]
    \centering
    \includegraphics[scale=0.2]{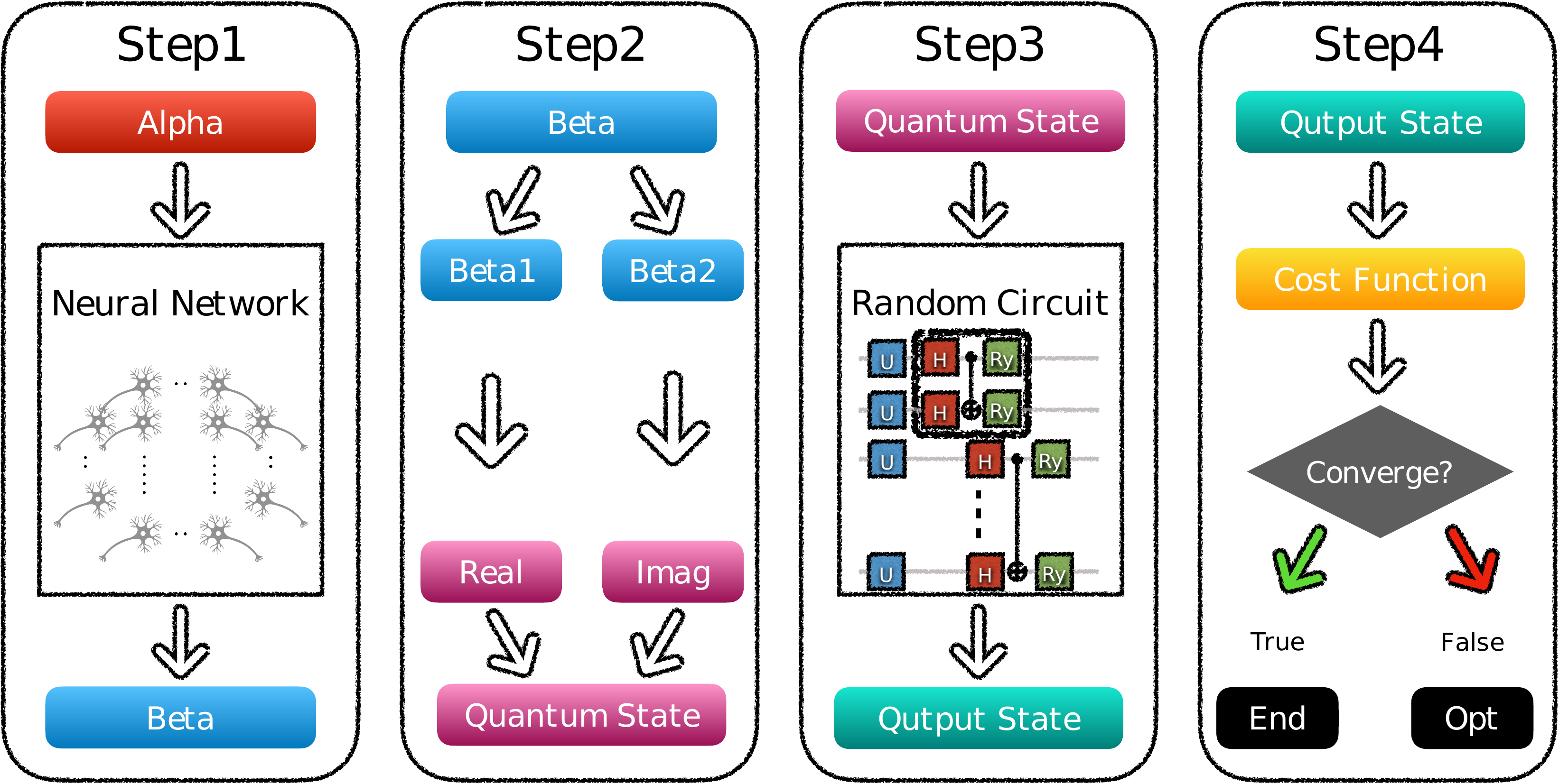}
    \caption{Illustration of NGQS models. In the NGQS model, the optimization step adjusts both $\alpha$ and the neural network, while the RQS model optimizes only $\alpha$. In the random quantum circuit, the box component corresponds to the defined block.}
    \label{fig:flow}
\end{figure*}

In the context of research on variational quantum algorithms, a significant issue known as barren plateaus (BPs) arises. To mitigate the BPs, various strategies have been proposed \cite{lqnncnn, isabp, llqnn}. Inspired by Ref.\cite{arnqs}, we introduce a strategy by using neural network generating quantum states (NGQS) to mitigate barren plateaus in variational quantum circuits. First, We extend the construction of Variational Quantum Circuits (VQCs) to facilitate a transition from a fixed configuration to a random configuration \cite{evqc}. To construct the random circuit, we first apply $R_x$, $R_y$ and $R_z$ gates on $N$ qubits. This operation is equivalent to applying a one-qubit unitary operation to each qubit, which can be decomposed as $U=R_x(\theta_i)R_y(\theta_j)R_z(\theta_k) $. Subsequently, we will calculate the depth $D$ of the quantum circuit and randomly apply a block to pairs of qubits, with the number of blocks corresponding to the circuit's depth. In each block, the Hadamard gate($H$) is first applied to the two selected qubits, followed by the application of a $CNOT$ gate to entangle them, and concluding with the application of the $R_y$ gate. The block can be expressed as $\prod\limits_{d=1}^D(\bigotimes_{j\in S_q(d)} H^{(j)}CNOT_dR_{y}^{(j)})$, where $S_q(d)$ denotes the two selected qubits. This approach leverages the unique features of quantum circuits, including rotation and entanglement, to better demonstrate the circuit's diversity and randomness. The entire circuit can be expressed as $\ket\psi = \bigotimes_{i}^{N}U_i\prod\limits_{d=1}^D(\bigotimes_{j\in S_q(d)} H^{(j)}CNOT_dR_{y}^{(j)})$. Second, we constructed three fully connected neural networks (FNNs) and one convolutional neural network (CNN) for generating quantum states, utilizing PReLU as the activation function, shown in the Table\ref{construction}. Initially, a vector $\alpha$ is randomly generated to serve as the input for the neural network. The output $\beta$ of the neural network is used to construct the quantum state. If the number of qubits is $N$, then the dimension of $\beta$ is required to be $2^{N+1}$. To align with the characteristics of the quantum state, we should normalize the data in $\beta$. We select the first $2^N$ terms of $\beta$ to represent the real components of the quantum state,  while the last $2^N$ terms correspond to the imaginary components. The selection of the real and imaginary components can be made arbitrarily, as the randomness of the vector will persist even after it has been processed by the neural network, unless a specific task is defined. Subsequently, the quantum state is applied to a randomly generated quantum circuit. The gradient is calculated using the state obtained after the quantum circuit and the cost function. The cost function is $C = 1 - \frac{1}{n}\sum_i Tr[(\ket{0}\bra{0}_i)\otimes\Romanbar{I}_{\overline{i}}\ket{\psi}]$, $i$ denotes the $i-th$ qubit, the cost function computes the probability that the final state is non-zero \cite{abpcdnn, evqc}. Then $\alpha$ and networks are updated using a gradient-based optimizer. We repeat the above process until the $C$ converges or the maximum number of iterations is achieved. We refer to the model that does not use neural networks as the Random Quantum State (RQS) model. All structures and processes are illustrated in Fig.\ref{fig:flow}
\begin{table*}[htbp]
 \centering \caption{The construction of neural networks}
\begin{tabular}{cccc}
    \hline
    FNN1 & FNN2 & FNN3 & CNN \\
    \hline
    Linear(4, 16) & Linear(4, 64) & Linear(4, 16) & Conv1d(1$^{*}$, 16) \\
    Linear(16, 2$^{N+1}$) & Linear(64, 2$^{N+1}$) & Linear(16, 32) &Conv1d(16, 32) \\
     &  & Linear(32, 2$^{N+1}$) & Conv1d(32, $\frac{2^{N+1}}{4}$+1)\\
    \hline
   \end{tabular}
   \begin{itemize}
       \footnotesize
       \item[*]The input or output channels of CNN. 
   \end{itemize}
   \label{construction}
    \end{table*}

\begin{figure*}[!htbp]
\centering
\subcaptionbox{Average iterations for 3 to 8 qubits\label{avergae}}{%
    \includegraphics[scale=0.34]{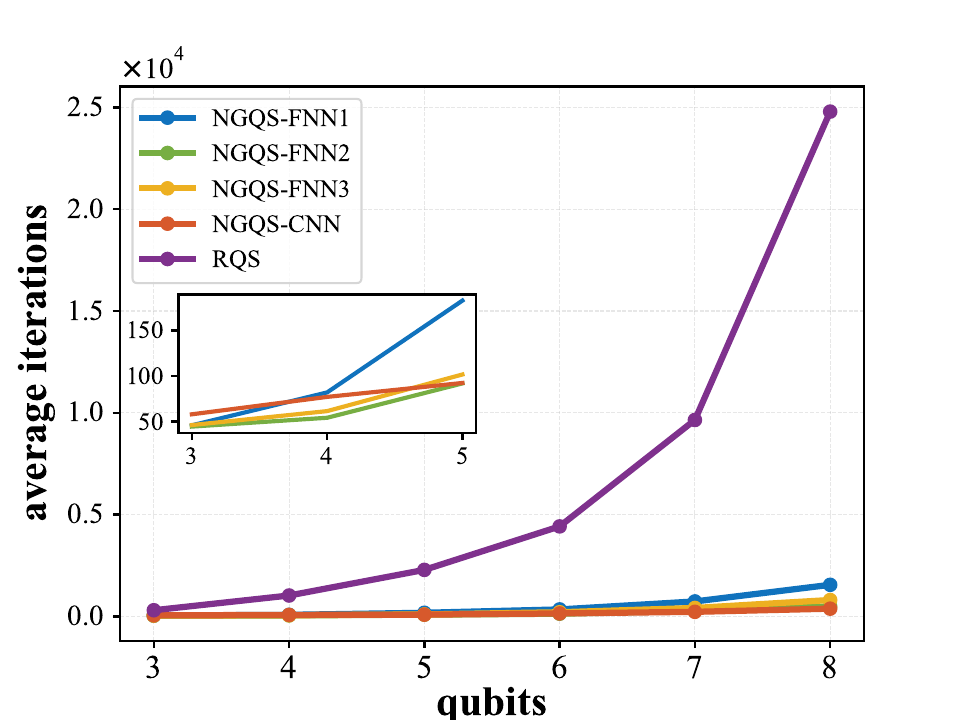}
}
\subcaptionbox{8qubits (Depth $N^2$)\label{N2}}{%
    \includegraphics[scale=0.34]{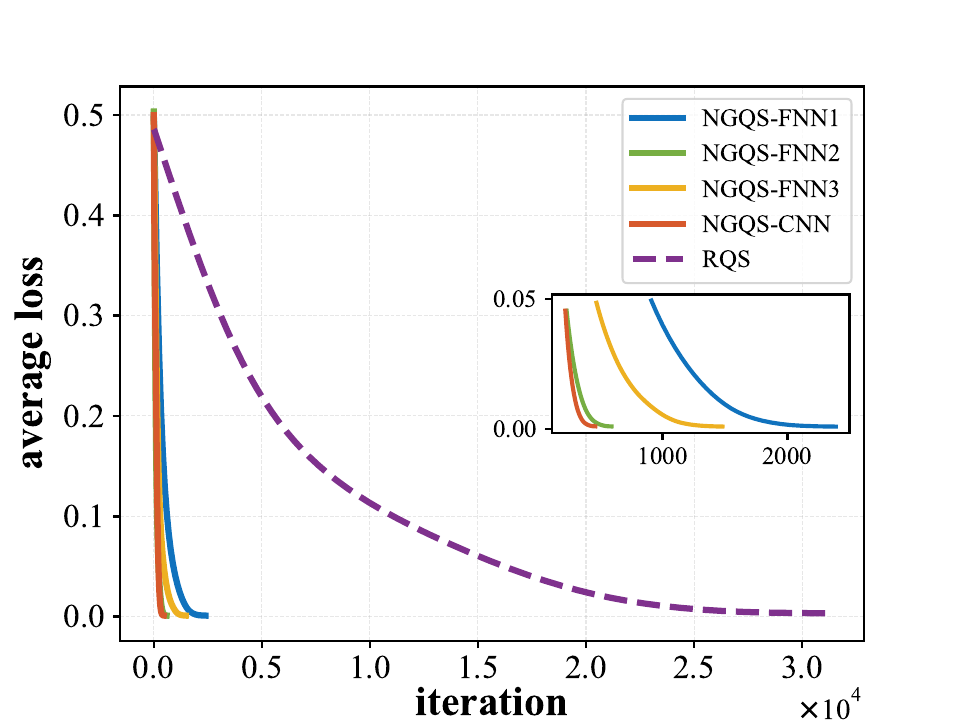}
}
\subcaptionbox{8 qubits (Depth $N^2\log(N)$)\label{N2logN}}{%
    \includegraphics[scale=0.34]{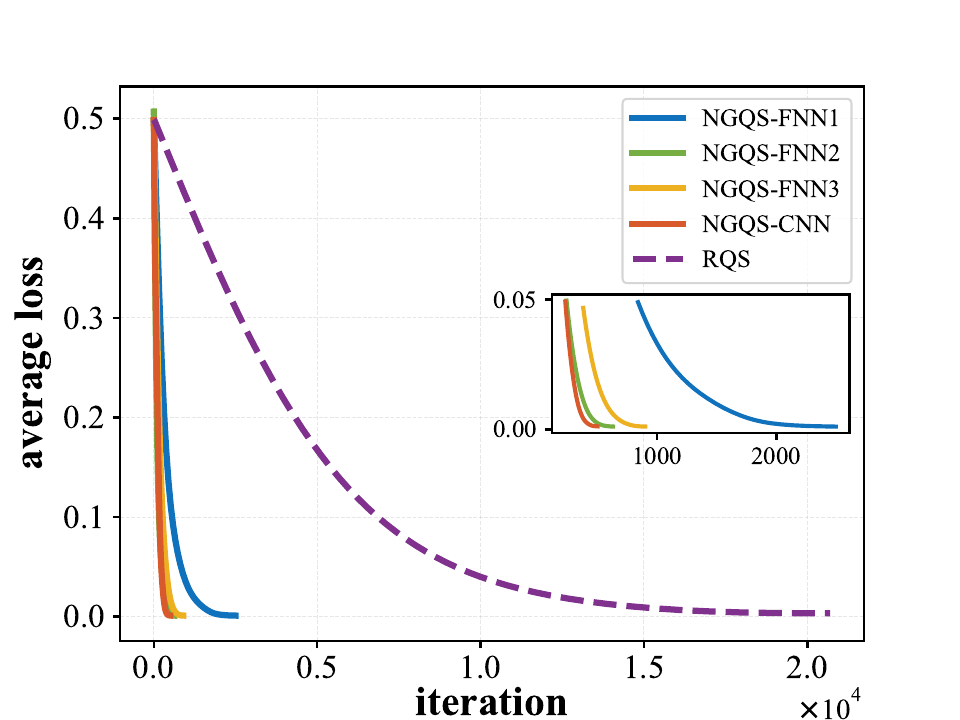}
}
\caption{Subfigure \textbf{(a)} depicting the average loss during training for ten models with different numbers of qubits; 
Subfigure \textbf{(b)} and Subfigure \textbf{(c)} depicting the average loss versus iterations during training at different depths.}
\label{interations_vs_loss}
\end{figure*}

\section{Discussion of the numerical result}
We now discuss our results in the context of NGQS and RQS. We apply stochastic gradient descent (SGD) with momentum as the optimization method. First, we set the depth of the quantum circuit to 30. For each model, we generate the corresponding random quantum circuit and the required circuit parameters based on the number of qubits and blocks. The circuit parameters are randomly selected in [0, 2$\pi$]. This process is repeated ten times. We collect the number of iterations for each trial and calculate the average. 

Fig.\ref{avergae} illustrates the average iterations required to achieve convergence for different number of qubits, with different lines representing distinct neural network architectures. As the number of qubits increases, the iterations required for the RQS model to converge rise explosively. This indicates that the RQS model, using the defined cost function, has suffered a barren plateau \cite{cf}. Although the iterations for the NGQS model also increase, this rise is much smaller compared to the RQS model, which required to achieve convergence is decreased by an order of magnitude. This indicates that the NGQS model can effectively mitigate the suffer from barren plateaus, regardless of the cost function. Moreover, CNN's performance is improving, followed by FNN2. Both CNN and FNN2 have more parameters than the other two models. We can infer that the number of parameters in a model influences its mitigation effect, with an increase in parameters resulting in improved performance. Next, we select $N^2$ and $N^2\log(N)$ to calculate the depth of the circuit. As shown in Figures \ref{N2} and \ref{N2logN}, the shaded regions represent the range between the maximum and minimum loss. In circuits of different depths, the RQS model is again effected by the barren plateau, while the NGQS model effectively mitigates this phenomenon.


\begin{thebibliography}{99} 

\bibitem{lqnncnn}G. Verdon, M. Broughton, J.R. McClean, K.J. Sung, R. Babbush, Z. Jiang, H. Neven, M. Mohseni, Learning to learn with
quantum neural networks via classical neural networks. arXiv:1907.05415 (2019)

\bibitem{isabp}E. Grant, L. Wossnig, M. Ostaszewski, M. Benedetti, An initialization strategy for addressing barren plateaus in
parametrized quantum circuits. Quantum \textbf{3}, 214 (2019)

\bibitem{llqnn} A. Skolik, J.R. McClean, M. Mohseni, P. van der Smagt, M. Leib, Layerwise learning for quantum neural networks. Quantum
Mach. Intell. \textbf{3}, 5 (2021)

\bibitem{arnqs}Lange, H., Van de Walle, A., Abedinnia, A. and Bohrdt, A., 2024. From architectures to applications: A review of neural quantum states. arXiv:2402.09402.

\bibitem{abpcdnn}L.Friedrich, J.Maziero, Avoiding barren plateaus with classical deep neural networks. Phys. Rev. A \textbf{106}(4), 042433 (2022)

\bibitem{evqc}Yi, Z., Liang, Y. and Situ, H., Enhancing variational quantum circuit training: an improved neural network approach for barren plateau mitigation. Physica Scripta, \textbf{100(8)}, p.086004, 2025.

\bibitem{cf}M. Cerezo, A. Sone, T. Volkoff, L. Cincio, and P. J. Coles, Cost function dependent barren plateaus in shallow parametrized quantum circuits, Nat. Commun. \textbf{12}, 1791 (2021).

\end{thebibliography}
\end{document}